\documentclass[preprint,preprintnumbers,amsmath,amssymb]{revtex4}

\usepackage{graphicx}
\usepackage{dcolumn}
\usepackage{bm}

\begin{document}

\preprint{Nuclei in the Cosmos IX, June 25-30 2006, CERN, Geneva, Switzerland}

\title{Present status of the KADoNiS database}

\author{I. Dillmann}\email{iris.dillmann@ik.fzk.de}
 \altaffiliation[also at ]{Departement f\"ur Physik und Astronomie, Universit\"at Basel.}
\author{R. Plag}
\author{M. Heil}
\author{F. K\"appeler}
 \affiliation{Institut f\"ur Kernphysik, Forschungszentrum Karlsruhe, Postfach 3640, D-76021 Karlsruhe,
        Germany}
\author{T. Rauscher}
 \affiliation{Departement Physik und Astronomie, Universit\"at Basel, Klingelbergstrasse 82, CH-4056 Basel,
        Switzerland}

\date{\today}

\begin{abstract}
The "Karlsruhe Astrophysical Database of Nucleosynthesis
in Stars" (KADoNiS) project is an online database for experimental
cross sections relevant to the $s$ process and $p$ process. It is
available under \emph{http://nuclear-astrophysics.fzk.de/kadonis}
and consists of two parts. Part 1 is an updated sequel to the
well-known Bao et al. compilations from 1987 and 2000, which is
online since April 2005. An extension of this $s$-process database
to $(n,p)$ and $(n,\alpha)$ cross sections at $kT$= 30 keV, as in
the f\-irst version of the Bao compilation, is planned. The second
part of KADoNiS is a $p$-process library, which includes all
available experimental data from $(p,\gamma)$, $(p,n)$,
$(\alpha,\gamma)$, $(\alpha,n)$, $(\alpha,\alpha)$, $(n,\alpha)$
and $(\gamma,n)$ reactions in or close to the respective Gamow
window. Despite the great number of reactions required for a
$p$-process reaction network, experimental data is still scarce
and up to now restricted to stable targets. Given here is a short
overview about the present status of the KADoNiS database.
\end{abstract}

\maketitle

\section{History of stellar neutron capture compilations}
The f\-irst collection of stellar neutron capture cross sections
was published in 1971 by Allen and co-workers \cite{alle71}. This
paper reviewed the role of neutron capture reactions in the
nucleosynthesis of heavy elements and presented also of a list of
recommended (experimental or semi-empirical) Maxwellian averaged
cross sections at $kT$= 30 keV (MACS30) for nuclei between carbon
and plutonium (6$\leq$Z$\leq$94). The idea of an experimental and
theoretical stellar neutron cross section database was picked up
again by Bao and K\"appeler \cite{bao87} for $s$-process studies.
This compilation published in 1987 included cross sections for
$(n,\gamma)$ reactions (between $^{12}$C and $^{209}$Bi), some
$(n,p)$ and $(n,\alpha)$ reactions ($^{33}$Se to $^{59}$Ni), and
also $(n,\gamma)$ and $(n,f)$ reactions for long-lived actinides.
A (limited) updated compilation was published in 1992 by Beer,
Voss and Winters \cite{BVW92}.

In the update of 2000 \cite{bao00} the Bao compilation from 1987
was extended to isotopes between $^{1}$H and $^{209}$Bi, and
listed -- like the original Allen paper -- also semi-empirical
re\-commended values for nuclides without experimental cross
section information. These estimated values are normalized cross
sections derived with the Hauser-Feshbach code NON-SMOKER
\cite{RaT01}, which account for known systematic def\-iciencies in
the nuclear input of the calculation. Additionally, the database
provided stellar enhancement factors and energy-dependent MACS for
energies between \emph{kT}= 5 keV and 100 keV.

\section{The updated big bang and $s$-process database}
The f\-irst update of the big bang and $s$-process database was
f\-inished in January 2006 \cite{DHK06}. In this stage, eight
theoretical cross sections ($^{74}$Se, $^{84}$Sr, $^{128-130}$Xe,
$^{147}$Pm, $^{151}$Sm, and $^{180}$Ta$^{m}$) were replaced with
new experimental values. Further 20 cross sections were updated by
the inclusion of new data. In order to make these updates
reproducible, a full list of updated cross sections with
references can be found on the KADoNiS homepage in the menu
section "Logbook". Here will be also a history of previous KADoNiS
versions.

The main future efforts in this part will be focussed on the
re-evaluation of semi-empirical cross sections and the
re-calculation of cross sections for isotopes, where a recent
change in physical properties (e.g. t$_{1/2}$, I$_\gamma$...)
leads to changes in already measured cross sections. Another
module will be the extension to $(n,p)$ and $(n,\alpha)$ cross
sections at $kT$= 30 keV, as in the f\-irst version of the Bao
compilation \cite{bao87}.

\section{The new $p$-process database}
The second part of KADoNiS is a collection of all experimental
reaction rates relevant for $p$-process studies. The f\-irst stage
is presented here and includes all available datasets of
$(\alpha,\gamma)$ (Table~\ref{tab:ag}) and $(p,\gamma)$
(Table~\ref{tab:pg}) reactions within the respective Gamow window
(see column E$_{Gamow}$).

\begin{table}[!ht]
\begin{center}
\begin{tabular}{cccccc}
\hline Isotope & Reaction & E$_{exp}$ $[MeV]$ & E$_{Gamow}$ $[MeV]$ & Reference & EXFOR Entry \\
\hline
$^{56}$Fe & $(\alpha,\gamma)$ & 3.90 - 6.50 & 3.21 - 6.86 & \cite{AKS79} & A0308 \\
$^{58}$Ni & $(\alpha,\gamma)$ & 4.90 - 6.10 & 3.40 - 7.18 & \cite{MSS64} & C0703 \\
$^{62}$Ni & $(\alpha,\gamma)$ & 5.10 - 8.60 & 3.40 - 7.19 & \cite{ZDE79} & C0669 \\
$^{64}$Ni & $(\alpha,\gamma)$ & 4.40 - 7.10 & 3.40 - 7.20 & \cite{ZDE79} & C0669 \\
$^{63}$Cu & $(\alpha,\gamma)$ & 5.90 - 8.70 & 3.49 - 7.34 & \cite{BNS05} & C1050 \\
$^{70}$Ge & $(\alpha,\gamma)$ & 5.05 - 7.80 & 3.78 - 7.80 & \cite{FKS96} & O0897 \\
$^{96}$Ru & $(\alpha,\gamma)$ & 7.03 - 10.56 & 4.85 - 9.51 & \cite{Rap02} & A0451 \\
$^{112}$Sn & $(\alpha,\gamma)$ & 8.30 - 9.97 & 5.32 - 10.30 & \cite{OMB02} & C0904 \\
$^{144}$Sm & $(\alpha,\gamma)$ & 10.50 - 13.40 & 6.26 - 11.78 & \cite{SFK98} & A0416 \\
\hline
\end{tabular}
\caption{List of $(\alpha,\gamma)$ reactions within the Gamow
window (E$_{Gamow}$), which are included in the $p$-process
library.} \label{tab:ag}
\end{center}
\end{table}

\begin{table}[!hp]
\begin{center}
\begin{scriptsize}
\begin{tabular}{cccccc}
\hline Isotope & Reaction & E$_{exp}$ $[MeV]$ & E$_{Gamow}$ $[MeV]$ & Reference & EXFOR Entry \\
\hline
$^{58}$Ni & $(p,\gamma)$ & 1.32 - 2.74 & 1.14 - 3.22 & \cite{KNE74} & A0696 \\
$^{58}$Ni & $(p,\gamma)$ & 1.00 - 4.91 & 1.14 - 3.22 & \cite{ChK80} & C0886 \\
$^{58}$Ni & $(p,\gamma)$ & 0.51 - 3.09 & 1.14 - 3.22 & \cite{KNE77} & A0048 \\
$^{58}$Ni & $(p,\gamma)$ & 1.14 - 4.09 & 1.14 - 3.22 & \cite{TMS85} & A0311 \\
$^{60}$Ni & $(p,\gamma)$ & 0.61 - 2.94 & 1.14 - 3.22 & \cite{KNE77} & A0048 \\
$^{61}$Ni & $(p,\gamma)$ & 1.11 - 2.94 & 1.14 - 3.22 & \cite{KNE77} & A0048 \\
$^{64}$Ni & $(p,\gamma)$ & 1.11 - 2.94 & 1.14 - 3.22 & \cite{SMA83} & A0198 \\
$^{63}$Cu & $(p,\gamma)$ & 1.11 - 4.69 & 1.17 - 3.29 & \cite{SMA83} & A0198 \\
$^{63}$Cu & $(p,\gamma)$ & 1.99 - 4.52 & 1.17 - 3.29 & \cite{Qia90} & C0739 \\
$^{65}$Cu & $(p,\gamma)$ & 1.03 - 3.22 & 1.17 - 3.29 & \cite{SMA83} & A0198 \\
$^{65}$Cu & $(p,\gamma)$ & 1.99 - 4.52 & 1.17 - 3.29 & \cite{Qia90} & C0739 \\
$^{64}$Zn & $(p,\gamma)$ & 1.47 - 2.73 & 1.21 - 3.35 & \cite{KNE77} & A0048 \\
$^{67}$Zn & $(p,\gamma)$ & 1.47 - 2.92 & 1.21 - 3.35 & \cite{KNE77} & A0048 \\
$^{68}$Zn & $(p,\gamma)$ & 1.67 - 4.97 & 1.21 - 3.35 & \cite{ESZ81} & C0650 \\
$^{74}$Se & $(p,\gamma)$ & 1.60 - 3.00 & 1.34 - 3.61 & \cite{KNE77} & A0048 \\
$^{74}$Se & $(p,\gamma)$ & 1.46 - 3.55 & 1.34 - 3.61 & \cite{GFS03} & O0849 \\
$^{76}$Se & $(p,\gamma)$ & 1.46 - 3.55 & 1.34 - 3.61 & \cite{GFS03} & O0849 \\
$^{77}$Se & $(p,\gamma)$ & 1.55 - 2.97 & 1.34 - 3.61 & \cite{KNE77} & A0048 \\
$^{84}$Sr & $(p,\gamma)$ & 1.67 - 2.96 & 1.47 - 3.85 & \cite{GSF01} & A0426 \\
$^{86}$Sr & $(p,\gamma)$ & 1.48 - 2.96 & 1.47 - 3.85 & \cite{GSF01} & A0426 \\
$^{87}$Sr & $(p,\gamma)$ & 1.58 - 2.96 & 1.47 - 3.85 & \cite{GSF01} & A0426 \\
$^{88}$Sr & $(p,\gamma)$ & 1.38 - 4.94 & 1.47 - 3.85 & \cite{GDK03} & O1054 \\
$^{89}$Y  & $(p,\gamma)$ & 1.76 - 4.83 & 1.51 - 3.91 & \cite{TKS04} & O1182 \\
$^{90}$Zr & $(p,\gamma)$ & 1.97 - 5.70 & 1.54 - 3.97 & \cite{LFH87} &  \\
$^{96}$Zr & $(p,\gamma)$ & 3.50 - 6.00 & 1.54 - 3.97 & \cite{CMB99} & C0556 \\
$^{93}$Nb & $(p,\gamma)$ & 1.42 - 4.80 & 1.57 - 4.03 & \cite{HST01} & O0918 \\
$^{92}$Mo & $(p,\gamma)$ & 1.48 - 3.00 & 1.60 - 4.08 & \cite{SaK97} & A0653 \\
$^{94}$Mo & $(p,\gamma)$ & 1.48 - 2.49 & 1.60 - 4.08 & \cite{SaK97} & A0653 \\
$^{95}$Mo & $(p,\gamma)$ & 1.70 - 3.00 & 1.60 - 4.08 & \cite{SaK97} & A0653 \\
$^{98}$Mo & $(p,\gamma)$ & 1.48 - 3.00 & 1.60 - 4.08 & \cite{SaK97} & A0653 \\
$^{96}$Ru & $(p,\gamma)$ & 1.65 - 3.37 & 1.66 - 4.20 & \cite{BSK98} & A0654 \\
$^{98}$Ru & $(p,\gamma)$ & 1.65 - 3.37 & 1.66 - 4.20 & \cite{BSK98} & A0654 \\
$^{99}$Ru & $(p,\gamma)$ & 1.46 - 3.37 & 1.66 - 4.20 & \cite{BSK98} & A0654 \\
$^{100}$Ru & $(p,\gamma)$ & 1.46 - 3.37 & 1.66 - 4.20 & \cite{BSK98} & A0654 \\
$^{104}$Ru & $(p,\gamma)$ & 1.65 - 3.37 & 1.66 - 4.20 & \cite{BSK98} & A0654 \\
$^{102}$Pd & $(p,\gamma)$ & 2.53 - 4.17 & 1.72 - 4.31 & \cite{OMB02} & C0904 \\
$^{112}$Sn & $(p,\gamma)$ & 3.00 - 8.50 & 1.84 - 4.52 & \cite{CMB99} & C0556 \\
$^{116}$Sn & $(p,\gamma)$ & 2.63 - 4.18 & 1.84 - 4.52 & \cite{OMB02} & C0904 \\
$^{119}$Sn & $(p,\gamma)$ & 2.80 - 6.00 & 1.84 - 4.52 & \cite{CMB99} & C0556 \\
\hline
\end{tabular}
\caption{List of $(p,\gamma)$ reactions within the Gamow window
(E$_{Gamow}$), which are included in the $p$-process library.}
\label{tab:pg}
\end{scriptsize}
\end{center}
\end{table}

These tables are listed under
\emph{http://nuclear-astrophysics.fzk.de/kadonis/pprocess},
including hyperlinks to the respective EXFOR f\-iles. In the
present stage (July 2006) we have included 39 $(p,\gamma)$
datasets of 32 isotopes between $^{58}$Ni and $^{119}$Sn, and 9
$(\alpha,\gamma)$ reactions. For these isotopes we have created
datasheets similar to those in the $s$-process database, which
include tabulated cross sections data, from which the respective
$S$ factor and reaction rate is calculated. As additional
information, a graphical plot of the dataset(s) with the location
of the Gamow window is given. A graphical comparison with
theoretical predictions from MOST \cite{most05} and NON-SMOKER
\cite{RaT01} will also be included.

Up to the end of 2006 we plan to extend this database also to
other available reactions, e.g. $(p,n)$, $(p,\alpha)$,
$(\alpha,n)$, $(\alpha,p)$, $(n,\alpha)$ and $(n,p)$ reactions. A
further step will be the inclusion of photodissociation rates, and
the calculation of those rates from $(n,\gamma)$ reactions via
detailed balance. Although KADoNiS is thought to be a "dynamic"
database, which is updated regularly, a paper version will be
published in 2007.

\begin{acknowledgments}
This work was supported by the Swiss National Science Foundation
Grants 2024-067428.01 and 2000-105328.
\end{acknowledgments}

\end{document}